\documentclass[aps,pra,reprint,superscriptaddress]{revtex4-2}
\usepackage{amsmath}
\usepackage{amssymb}
\usepackage{graphicx}
\usepackage{dcolumn}
\usepackage{bm}
\usepackage{color}
\begin{document}

\title{Noise spectroscopy in Rydberg atomic ensemble}

\author{Jun He$^{1,2,3*}$, Qiang Liu$^1$, Ze Yang$^1$, Qiqi Niu$^1$, Xiaojuan Ban$^1$, Yijie Du$^3$, Junmin Wang$^{1,2**}$ \\ {\it 1 State Key Laboratory of Quantum Optics and Quantum Optics Devices, and Institute of
Opto-Electronics, Shanxi University, Tai Yuan 030006, People’s Republic of China.} \\
{\it 2 Collaborative Innovation Center of Extreme Optics, Shanxi University, Tai Yuan 030006, People’s Republic of China.} \\
{\it 3 Beijing Academy of Quantum Information Sciences, Beijing 100193, China.} \\
*hejun@sxu.edu.cn , ** wwjjmm@sxu.edu.cn}

\date{\today}

\begin{abstract}
The phase noise of incident light fields can be converted into amplitude noise via absorption and dispersion effects in media under electromagnetically induced transparency. The conversion process is sensitive to two-photon detuning. This effect can be used to perform a novel type of Rydberg spectroscopy. We present experimental measurements of the phase noise spectrum (PNS) with Rydberg atoms in a cesium vapor cell. Furthermore, to explore microwave field measurements, a modulation-enhanced PNS approach is proposed.
\end{abstract}

\maketitle
\section{INTRODUCTION}
Recent developments in quantum manipulation offer the possibility of performing field sensing using a pure quantum system. With regard to technical requirements for sensors in such systems, neutral atoms are convenient. Alkali metal atoms contained in chip-scale atomic vapor have been used in devices, including atomic clocks, atomic magnetometers, and atomic gyroscopes [1–3]. In particular, Rydberg atoms display exaggerated electrical characteristics. Transition energies between pairs of nearby Rydberg states are distributed over a wide range of frequencies, from megahertz to terahertz, that offer promising advantages when using these states for wideband electric field sensing [4, 5].

Optical transition frequencies are typically several hundreds of terahertz. Higher interaction frequencies strongly suppress the participation of the thermal population between energy levels of specific transitions. Atomic sensor sensitivity depends on its spectral resolution [4–6]. At lower frequency range, the signal-to-noise ratio (SNR) for spectroscopy is limited by classical sources of mechanical and audio noise. At higher frequencies, the sensitivity is limited by the quantum noise of photons and atomic ensembles [7–12]. Rydberg-atom sensing relies on an electromagnetically induced transparency (EIT)-based method, in which various laser fields are required to perform internal quantum state manipulations. EIT signal incur inhomogeneous broadening under high-intensity field coupling conditions [10]. To improve sensing sensitivity, the EIT parameters are controlled to operate under weak excitation conditions. Noise in the intensity of laser beams may be suppressed through a combination of averaging and differential detection [11, 12]. However, phase noise of the laser beams remains strong and introduces measurable noise. A phase-amplitude transformation has been observed in EIT systems, and studies of quantum mechanical treatments for both atomic and field fluctuations have been presented in the literature [13, 14]. Zhang and colleagues investigated noise property of delayed light in EIT, in which conjugate phase and amplitude quadratures were demonstrated experimentally [15]. Xiao and colleagues demonstrated resonance suppression in EIT media during the conversion of phase noise into intensity noise of a laser beam [16]. Li and colleagues demonstrated using a velocity-selective optical pumping technique that the resolution of the amplitude noise spectrum in cesium atomic vapor is enhanced by narrowing the absorption [17].

When the Rabi frequencies of the incident light fields are lower than the atomic spontaneous decay rate, including those of laser beams and microwave (MW) fields, the phase noise–amplitude noise (PN–AN) plays a significant role in EIT processes. The Rydberg EIT system is driven by probe and coupling fields in a ladder-type configuration. The sum of the probe and coupling frequencies determines the two-photon detuning. From theory and experiments, Gea-Banacloche and colleagues demonstrated a laser linewidth effect in Doppler-broadened EIT media [18]. Lü and colleagues experimentally studied EIT transmissions influenced by laser frequency or phase noise in the driving fields [19]. Kim and colleagues studied the effect of frequency noise on the decoherence rate of Rydberg EIT systems [20]. These noise or fluctuations randomly introduce a two-photon detuning of the system, resulting in an attenuation of the probe field. Simons and colleagues performed experiments measuring the radio-frequency E-field strengths in the presence of white Gaussian noise [21]. Jing and colleagues demonstrated a MW electric field sensor using Rydberg atoms of cesium [6]. The frequency or phase noise of laser beams were suppressed by locking these lasers to an ultralow-expansion glass cavity. However, deformations of the optical components and optical fibers on nanometer scales introduced additional phase fluctuations in the laser beams, and long-range transmissions of the MW field produce additional phase noise through stochastic scattering and reflection. Additional sources of phase noise in the laser beams or MW field would induce PN–AN conversions during EIT spectroscopy, which plays a significant role in Rydberg atom-based MW sensing.

The phase noise of incident light fields can be converted into amplitude noise of the probe signal via absorption and dispersion effects in Rydberg EIT media. The PN–AN conversion signal intensity is dependent on two-photon frequency detuning. This effect is used to perform a novel type of Rydberg spectroscopy. In this paper, we present a numerical simulation of PN-AN conversion in EIT media and report experimental demonstrations using Rydberg atoms in a cesium vapor cell. We propose a modulated phase noise spectral technique that extends the spectral analyzing frequency to the megahertz band.
\section{EXPERIMENT}
\begin{figure*}[htbp!]
\centering
\includegraphics[scale=0.42]{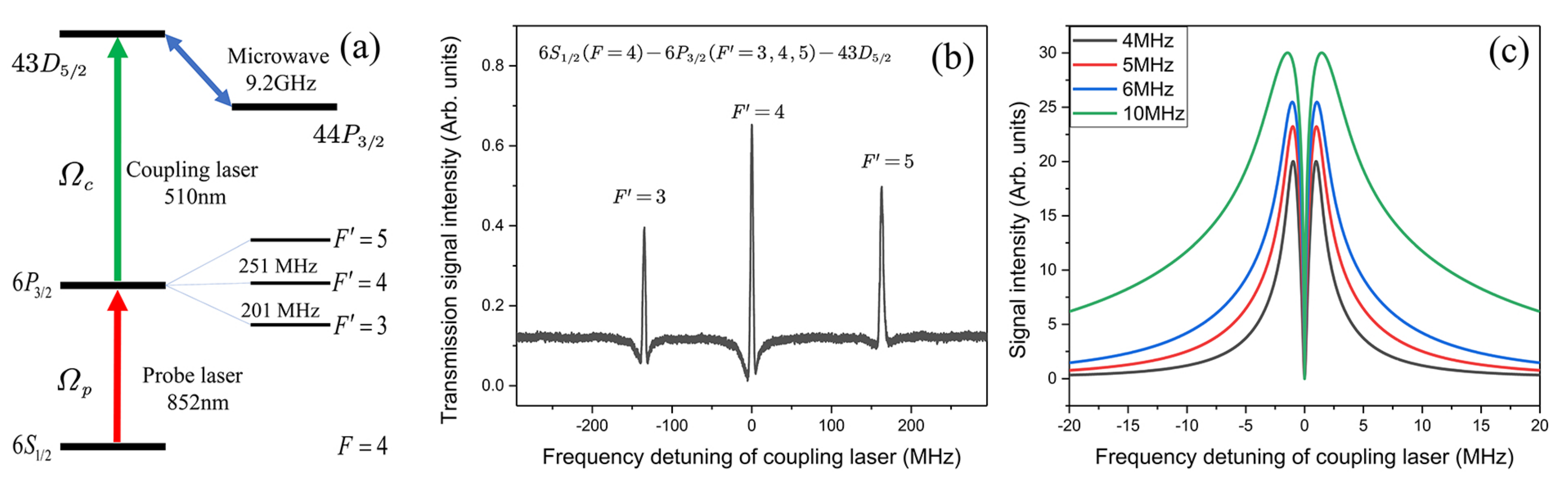}
\caption{\label{Fig_1}Schematics of the ladder-type system and its spectral characteristics: (a) Energy level scheme of the ladder-type EIT of the cesium atom; (b) EIT spectroscopy. All hyperfine-transition EITs of the intermediate states are observed because of velocity-selective effects in the room-temperature atomic vapor cell. The Rabi frequencies of the probe laser and the coupling laser are $\sim$4 MHz. The typical linewidths of the coupling and probe laser beams are $\sim$5 MHz and 0.1 MHz, respectively; (c) Converted signal intensity of phase noise versus the coupling laser frequency detuning. The Rabi frequency of the coupling laser beam has increased from 4 MHz to 10 MHz. The power of the probe laser beam is approximately four orders of magnitude weaker than that of the coupling laser beam. The spontaneous decay rates of the 6P$_{3/2}$ state and the 43D$_{5/2}$ state are $\sim$5.2 MHz and $\sim$1 kHz, respectively. The analyzing frequency used for numerical simulations is 0.9 MHz.}
\end{figure*}

We follow the approaches given in [15,16] to provide a qualitative theoretical picture of the phase noise spectrum (PNS) of incident light fields in a Rydberg atomic ensemble. The system may be simplified as a four-level model, as shown in Fig. 1(a). The weak probe beam of 852 nm laser couples levels $6S_{1/2}$ and $6P_{3/2}$, and the strong coupling beam of 510 nm laser couples levels $6P_{3/2}$ and $43D_{5/2}$. The Rabi frequencies of probe and coupling beams are denoted by $\omega_p$ and $\omega_c$, respectively. Our ladder-type EIT signal is shown in Fig. 1(b) and theoretical results for the PNS of the output signal are shown in Fig. 1(c). The EIT and PNS signals are obtained by scanning the coupling laser beam frequency across the upper transition while fixing the probe laser beam frequency. The PNS signals comprise three contributions: the intensity noise of the incident light beams, the Langevin noise arising from the random decay processes of the atoms, and the converted phase noise of the incident light beams. The first two of these contributions are very weak and hence neglected here. The third source contributes to the output signal as a result of the PN-AN conversion. In general, the amplitude noise spectrum includes four peaks; we only present here the simulation results using experiment dependent parameters. The typical linewidths of the coupling and probe laser beams are $\sim$5 MHz and $\sim$0.1 MHz, respectively. The spontaneous decay rates of the 6P$_{3/2}$ state and the 43D$_{5/2}$ state are $\sim$5.2 MHz and $\sim$1 kHz. The Rabi frequency of the coupling laser beam increases from 4 MHz to 10 MHz. The power of coupling laser beam is $\sim$200 mW and the power of probe laser beam $\sim$0.01 mW. The analyzing frequency of PNS output signal used for numerical simulations is 0.9 MHz. The PNS signal is dependent on the laser beams detuning. When the frequencies of the probe and coupling laser beams satisfy two-photon resonance conditions, the PN-AN conversion is then strongly suppressed and the PNS signal is much weak. The spectral profiles are narrow and spectral centers are insensitive to the driving fields intensity. The signal intensity increases with increasing power of coupling laser beams which is represented by the Rabi frequency, as shown in Fig. 1 (c) denoted by colorful lines. 
\begin{figure}
\centering
\includegraphics[scale=0.4]{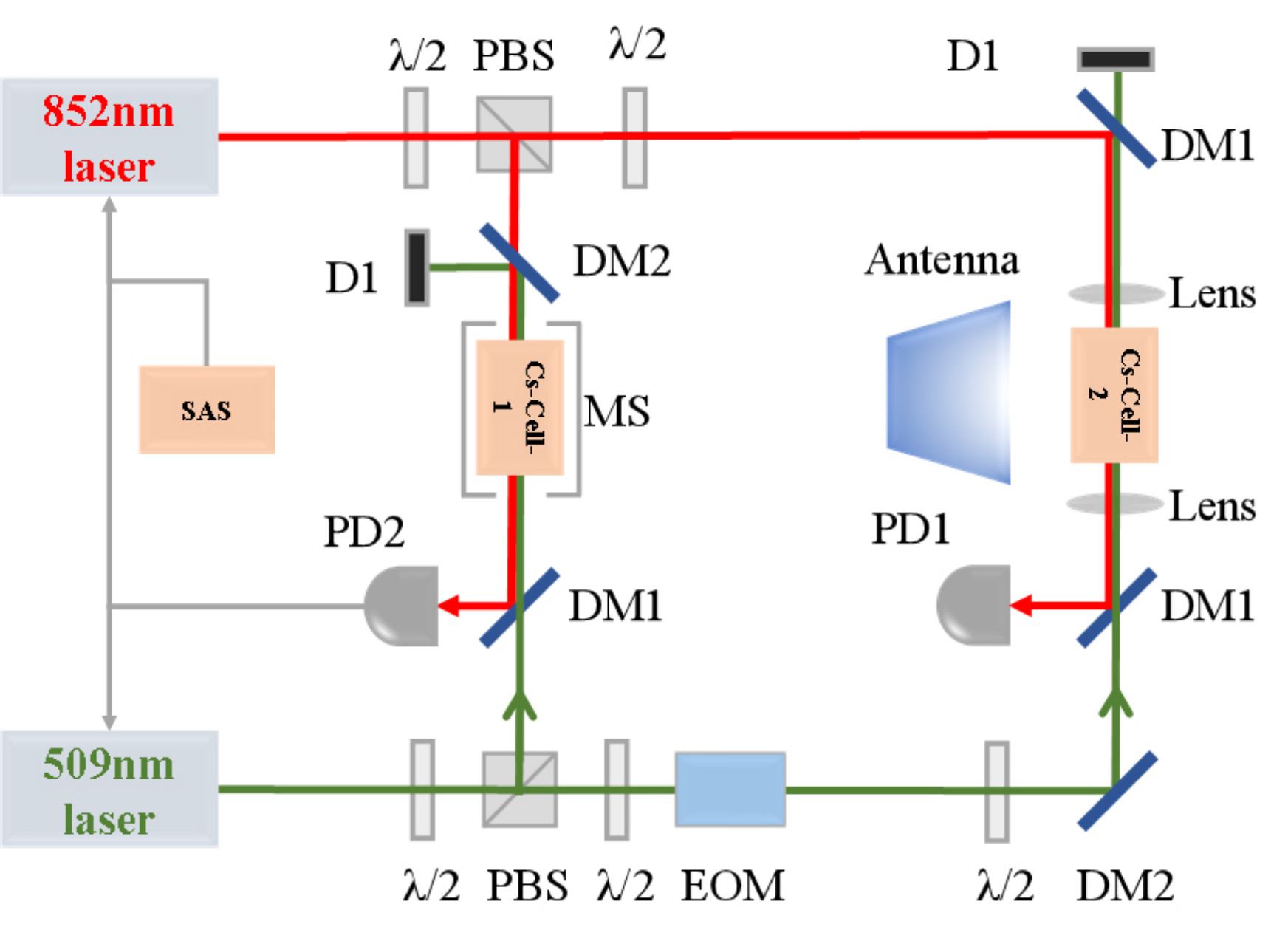}
\caption{\label{Fig_1}Schematic of the experimental apparatus. Both beams of 852 nm and 510 nm are overlapped in the cesium atomic vapor Cs-cell-1 and Cs-cell-2 using a counterpropagating configuration. The 852 nm laser frequency is stabilized to the hyperfine transition via saturation absorption spectroscopy (SAS). The 510 nm laser frequency is stabilized through Rydberg atom EIT spectroscopy in the Cs-cell-1. The EIT and PNS signals are performed in Cs-cell-2. Antenna: transforms MW signals from a signal generator to an electromagnetic wave in free space; $\lambda/2$: half-wave plate; PBS: polarizing beam splitter cube; EOM: electro-optic modulator; MS: magnetic shielding; PD: photodiode detector; DM1: 852 nm high-reflectivity (HR) and 510 nm high-transmissivity (HT) dichroic mirror; DM2: 852-nm HT and 510 nm HR dichroic mirror; D1: optical dump.}
\end{figure}

A schematic diagram of the experimental apparatus is shown in Fig. 2. An 852 nm external cavity diode laser (ECDL) with a typical linewidth of $\sim$ 0.1 MHz is used as a probe laser. The output optical power of a 1018 nm ECDL is amplified up to 5 W using an ytterbium-doped fiber amplifier and is then frequency-doubled in a periodically poled lithium niobate crystal to produce a 510 nm laser. These 852 nm and 510 nm beams are then overlapped in two cesium atomic vapor cells (Cs-cell-1 and Cs-cell-2) using a counterpropagating configuration. The two cells have a length of $\sim$2.5 cm to match the Rayleigh lengths of the focused beams, specifically, a $\sim$ 500 $\mu$m waist for the 510 nm beam and a $\sim$ 600 $\mu$m waist for the 852 nm beam. The phase-type electro-optic modulator (EOM) is used to introduce phase noise into laser beams. An antenna is used to transform MW signals from the signal generator into an electromagnetic wave propagating in free space. The transmission signal of 852 nm probe beam is monitored by photodiode detector. The 852 nm laser frequency is stabilized to the hyperfine transition via saturation absorption spectroscopy (SAS). The 510 nm laser beam is frequency-stabilized through Rydberg atom EIT spectroscopy in Cs-cell-1. The EIT and PNS signals are performed in Cs-cell-2. The EIT and PNS spectroscopy is performed by scanning the frequency detuning of the coupling laser beam with fixed frequency of probe laser beam.
\section{RESULTS AND DISCUSSION}
\subsection{Phase noise spectrum}
Figure 3(a) shows the intensity fluctuations of the probe signal versus the coupling laser frequency detuning in the time domain. The power of both probe and coupling beams have been stabilized using an opto-electrical feedback system. The probe laser beam is resonant with the transition $6S_{1/2}(F=4) \rightarrow 6P_{3/2}(F’=5)$, whereas frequency detuning of the coupling laser is turned relative to the transitions of the $6P_{3/2}(F’=5)$ state and the state $43D_{5/2}$. When the frequency detunings of coupling beam are $\pm$ 200 MHz, the measured signal fluctuations are dominated by the intensity noise of the probe laser beam. When both probe and coupling beams are near resonant, a strong intensity fluctuation occurs. In addition to laser beams noise intensity, there is another PN–AN noise source arising from a steep dispersion under EIT conditions. Figure 3(b) shows the power spectra of the probe signal for various coupling laser beam detunings. The spectral crossover point occurs at approximately 1 MHz for the different laser frequency detunings. For the typical analyzing frequency of 0.5 MHz, the measured noise intensity at near resonant frequency is approximately 5 dB higher than that measured with a +200 MHz frequency detuning.
\begin{figure*}
\centering
\includegraphics[scale=0.5]{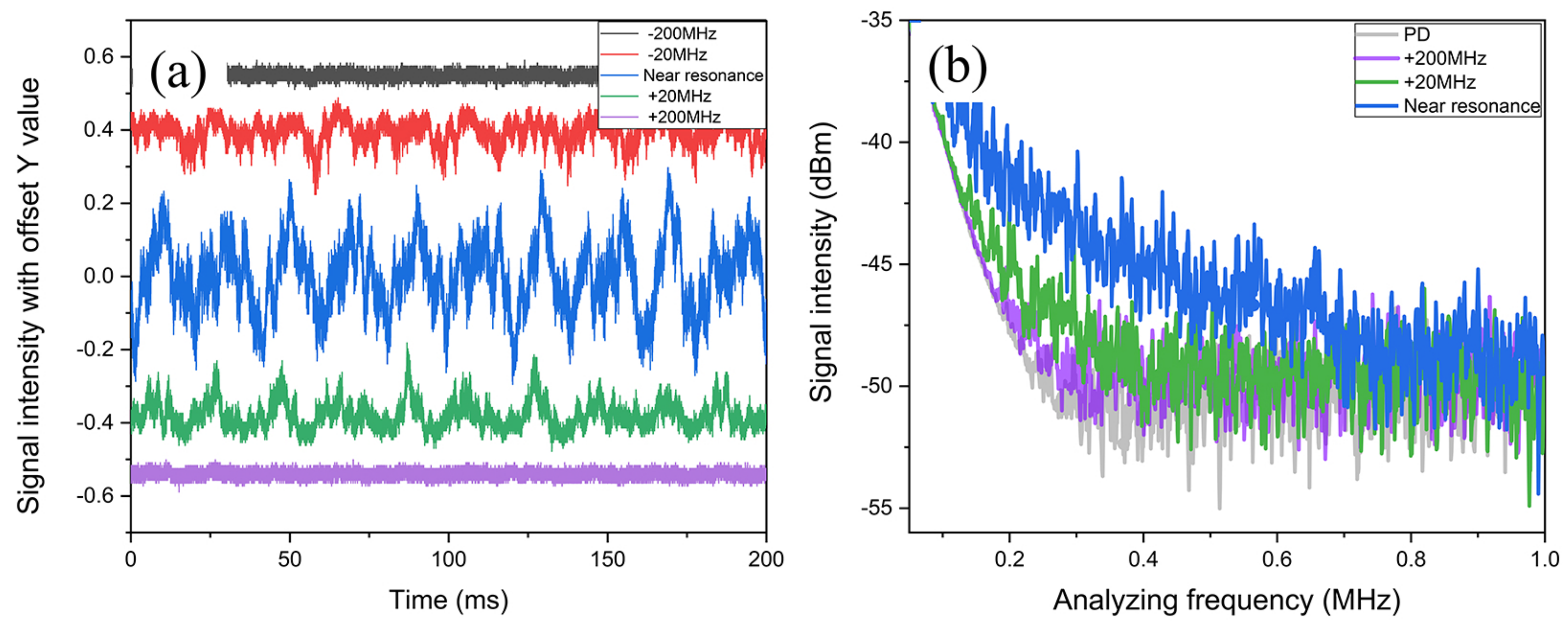}
\caption{Intensity fluctuations of the EIT transmission signals versus coupling laser beam frequency detunings: (a) Intensity fluctuations of the probe signals in the time domain. The signals are shifted by Y offsets, and the amplitudes of signals have not been scaled; When the frequency detunings are $\pm$200 MHz, the signal amplitudes are less than $\sim$13.5\% of that of the near resonant frequency. (b) Power spectra of the probe signals in the frequency domain. The spectral intensity at the near resonant condition is more than 5 dB greater in magnitude than that of the 200 MHz frequency detuning EIT. The spectral intensity depends on analyzing frequency. For the measurement frequency range 0.05 MHz-1 MHz, the resolution and video bandwidths are 100 kHz and 10 kHz, respectively. The electric dark noise is -53 dBm. All noise traces are averaged 16 times.
}
\end{figure*}

The PN–AN conversion is used to implement the Rydberg atom spectroscopy technique. The spectral signals are obtained by scanning the frequency detuning of the coupling laser while locking the probe laser frequency to the transition $6S_{1/2}(F=4) \rightarrow 6P_{3/2}(F’=4)$.The PNS traces (Fig. 4, blue and red curves) are recorded using a spectrum analyzer operating in the zero-span mode. The analyzing frequency of spectrum analyzer is of 0.9 MHz. The resolution bandwidth (RBW) and video bandwidth (VBW) are 100 kHz and 10 kHz, respectively. For comparisons, the EIT signal traces (Fig. 4, black curve) are recorded using an oscilloscope. These PNS spectral responses represent a multipeak structure with narrow central linewidth. The PNS signal intensity is dependent on the analyzing frequency, decreasing rapidly over the range from 0.3 MHz to 1.5 MHz, as shown in Fig. 5(a). Here, we propose a phase modulated PNS to improve the spectral SNR. The phase modulation is applied to the coupling laser beam by using a phase-type EOM, where the modulated frequency is of 1 MHz and the modulated bandwidth is of less than 1 Hz. A radio-frequency synthesizer, which was stabilized using a table-top rubidium atomic clock, is used to drive the EOM. The modulated PNS signals represent a significantly high SNR, as shown in Fig. 5(b). The modulated PNS can be used to extend the analyzing frequency range. The spectral distinguishable SNR is achieved at analyzing frequency of 1.2 MHz denoted by purple line in Fig. 5(b). The experimental analyzing frequency is varied from 0.3 MHz to 1.5 MHz, where the minimum RBW and the minimum VBW are 10 Hz and 1Hz, respectively. We have demonstrated in experiments that the PNS analyzing frequency can be extended from 100 kHz to $\sim$50 MHz, where the maximum analyzing frequency is limited to the EIT responding band. The signals with high analyzing frequency are achieved by strongly increasing the intensity of phase noise and intensity of the probe and coupling laser beams, which results in a strong spectral broadening.

\begin{figure}
\centering
\includegraphics[scale=0.17]{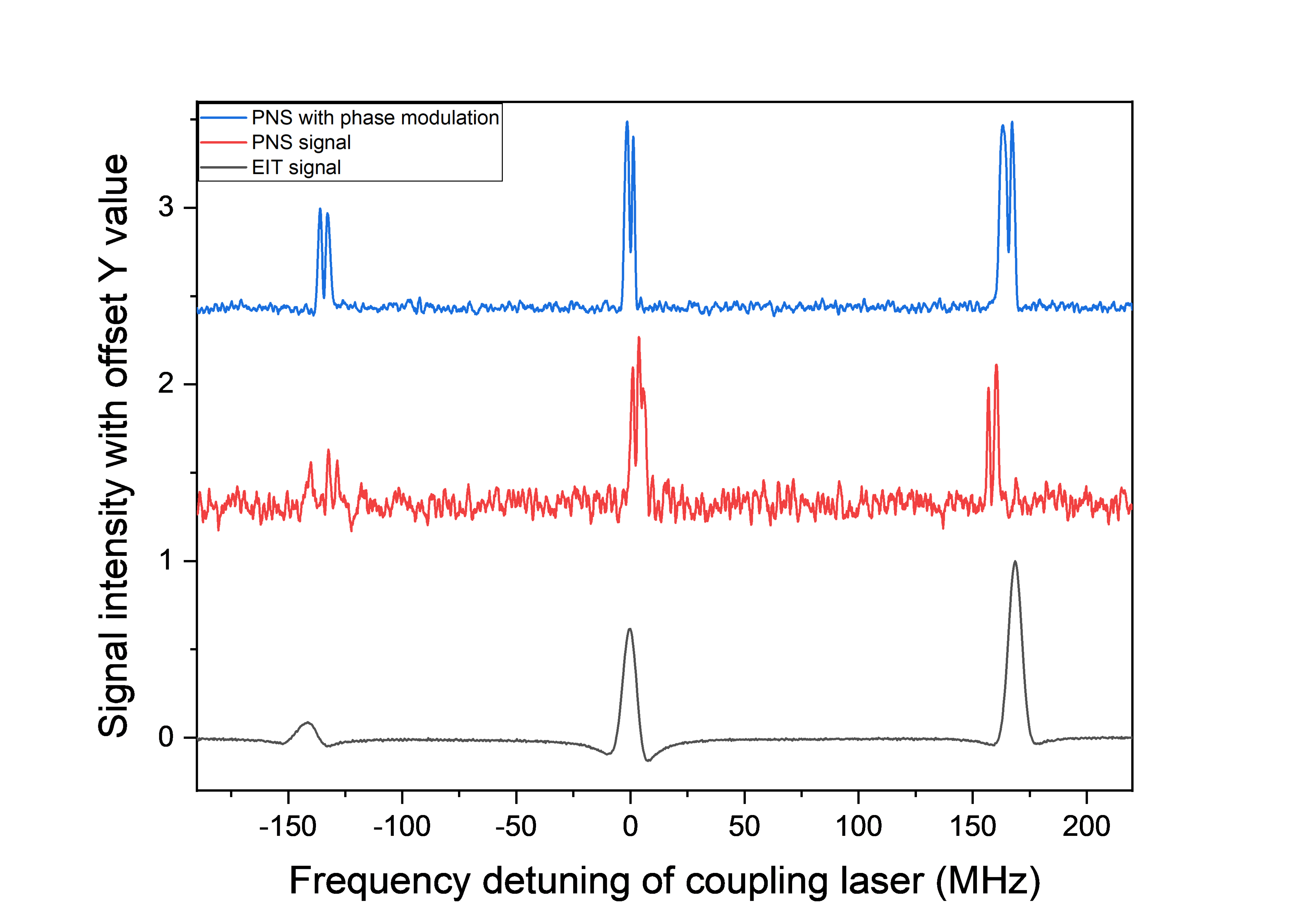}
\caption{Comparison of EIT, PNS, and modulated PNS. The spectral signals are monitored by scanning the coupling laser frequency while locking the probe laser frequency. These PNS spectral represent a multipeak structure with narrow central linewidth when the probe and coupling laser beam are in two-photon resonance. The line shape for modulated PNS is the same as that for PNS. All hyperfine transitions of the intermediate states are observed because of velocity-selective effects within the room-temperature atomic cell. The modulation signal is applied to the coupling laser beam by using an EOM, where the modulated frequency is of 1 MHz and the modulated bandwidth is of less than 1 Hz. The spectrum analyzer is set to the zero-span mode. The analyzing frequency is 0.9 MHz. The RBW and VBW are 100 kHz and 10 kHz, respectively.} 
\end{figure}
\begin{figure*}[htbp!]
\centering
\includegraphics[scale=0.35]{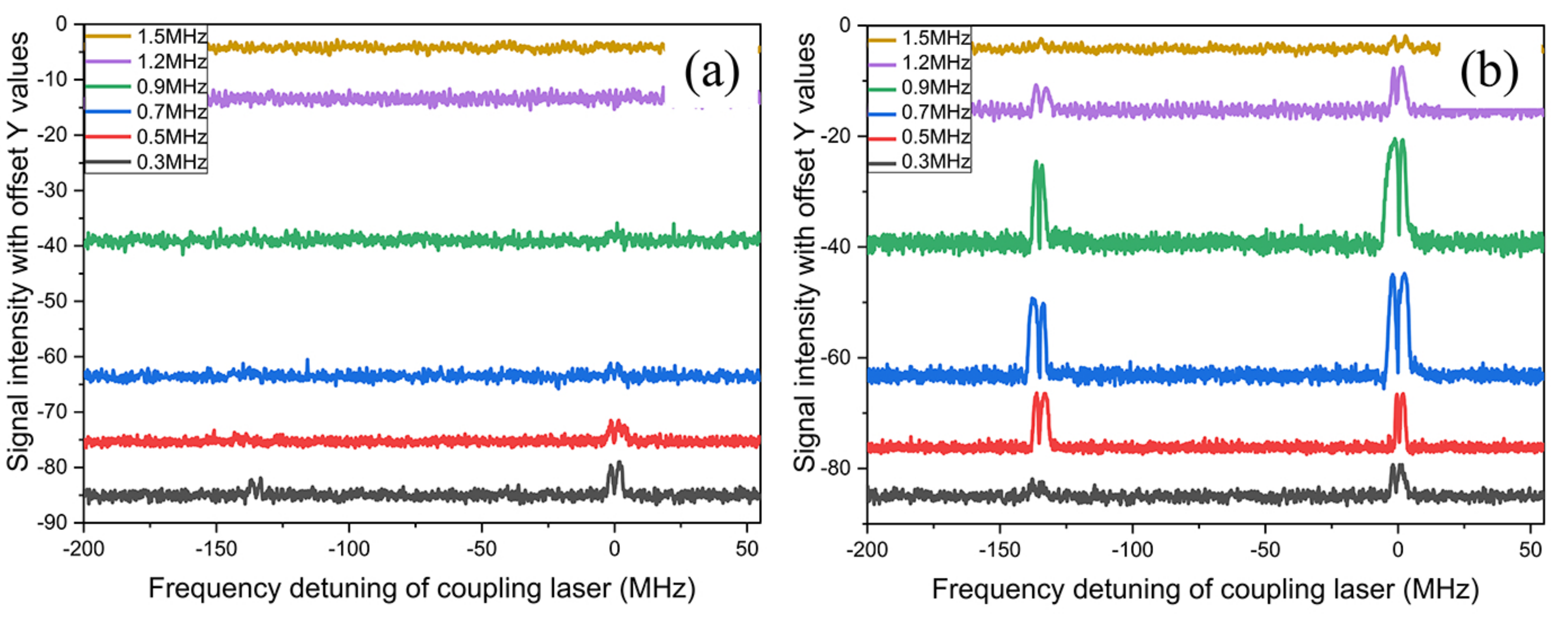}
\caption{Comparison of signal intensity of (a) PNS with (b) modulated PNS for various analyzing frequencies. The frequency of the modulation signal is 1 MHz. The spectrum analyzer operates in the zero-span mode with the analyzing frequencies varying from 0.3 MHz to 1.5 MHz. The RBW and VBW are changed with analyzing frequencies. The linewidth broadening for the modulated PNS is dominated by power broadening of the laser beams.} 
\end{figure*}

\subsection{Microwave measurements}
We use the modulated PNS to perform MW measurements. When the resonant MW field couples to the Rydberg states of $43D_{5/2}\rightarrow 44P_{3/2}$, the EIT signal splits through the Autler–Townes (AT) effect in a four-level atomic system. This splitting is proportional to the applied MW E-field amplitude. Figure 6 shows the spectral splitting of the EIT and the PNS spectrum when the MW field is switched on and off. By measuring the width of this splitting, the MW E-field strength can be obtained using the formula $|E|=2\pi \delta f\bar{h}/\mu$; here, $\bar{h}$ denotes the reduced Planck constant, $\mu$ is the atomic electric dipole moment of the MW transition, and $\delta f$ is the measured EIT-AT splitting. The MW E-field strength that is measured in this EIT-AT splitting method derives from the International System of Units (SI) which links it with Planck’s constant.

In our experiments, the transition frequency for the MW coupled states is 9.2 GHz, and the transition dipole moment is $\mu=3356 ea_0$, with e the electronic charge, and $a_0$, the Bohr radius. The Rabi frequencies of the coupling and probe laser beams are $\sim$3 MHz and $\sim$5 MHz, respectively. The modulated frequency of coupling laser beam is of 1 MHz. For the EIT spectrum, the measured frequency splitting is approximately $\delta f=(17\pm 3)MHz$, for which the corresponding electric field amplitude is approximately 5.2 mV/cm. For the PNS spectrum, the measured frequency splitting is approximately $\delta f=(15\pm 1)MHz$, the corresponding electric field amplitude being approximately 4.6 mV/cm.
\begin{figure*}
\centering
\includegraphics[scale=0.4]{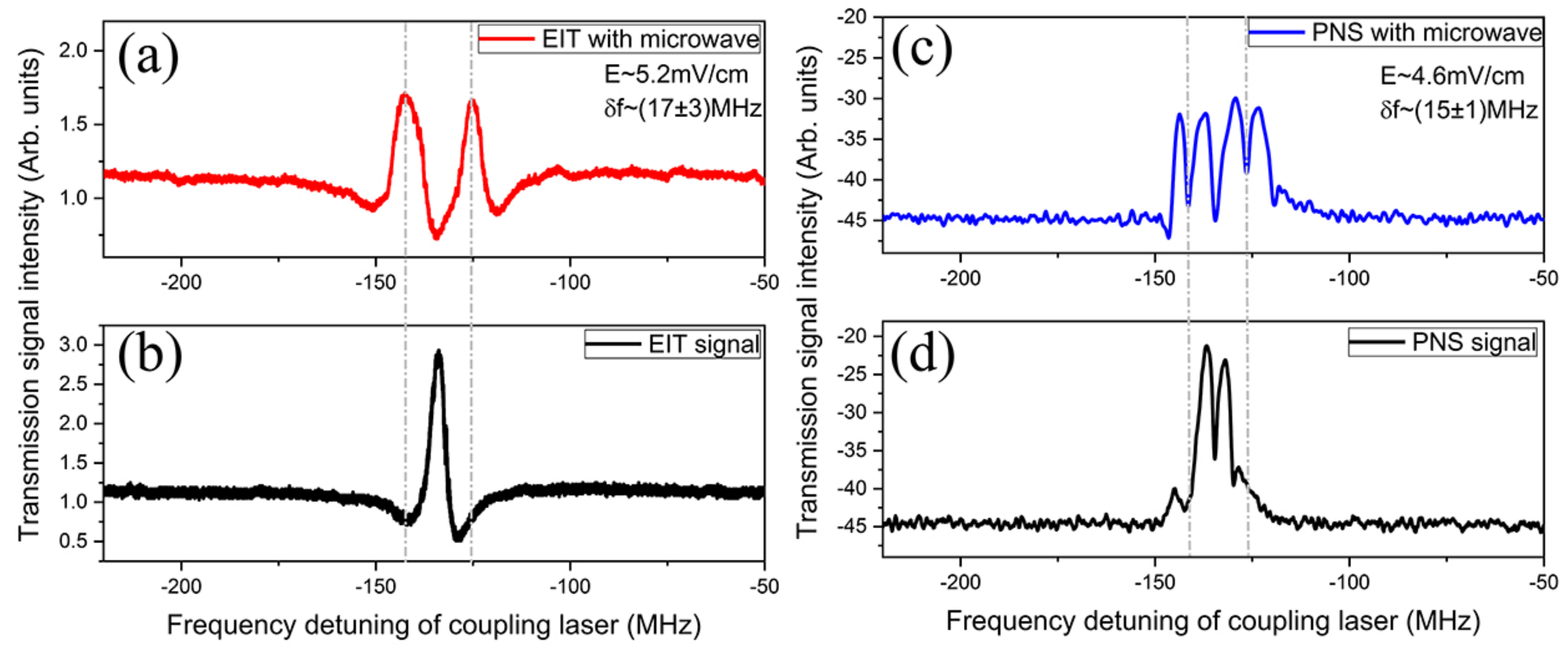}
\caption{(a, b) EIT and (c, d) PNS signals with and without MW fields. The MW transition causes splitting of the EIT and PNS via the AT effect. For the EIT spectrum, the measured frequency splitting is approximately $\delta f=(17\pm 3)MHz$, the corresponding electric field amplitude being approximately 5.2 mV/cm. For the PNS spectrum, the measured frequency splitting is approximately $\delta f=(15\pm 1)MHz$, the corresponding electric field amplitude being approximately 4.6 mV/cm.}
\end{figure*}

\begin{figure*}
\centering
\includegraphics[scale=0.4]{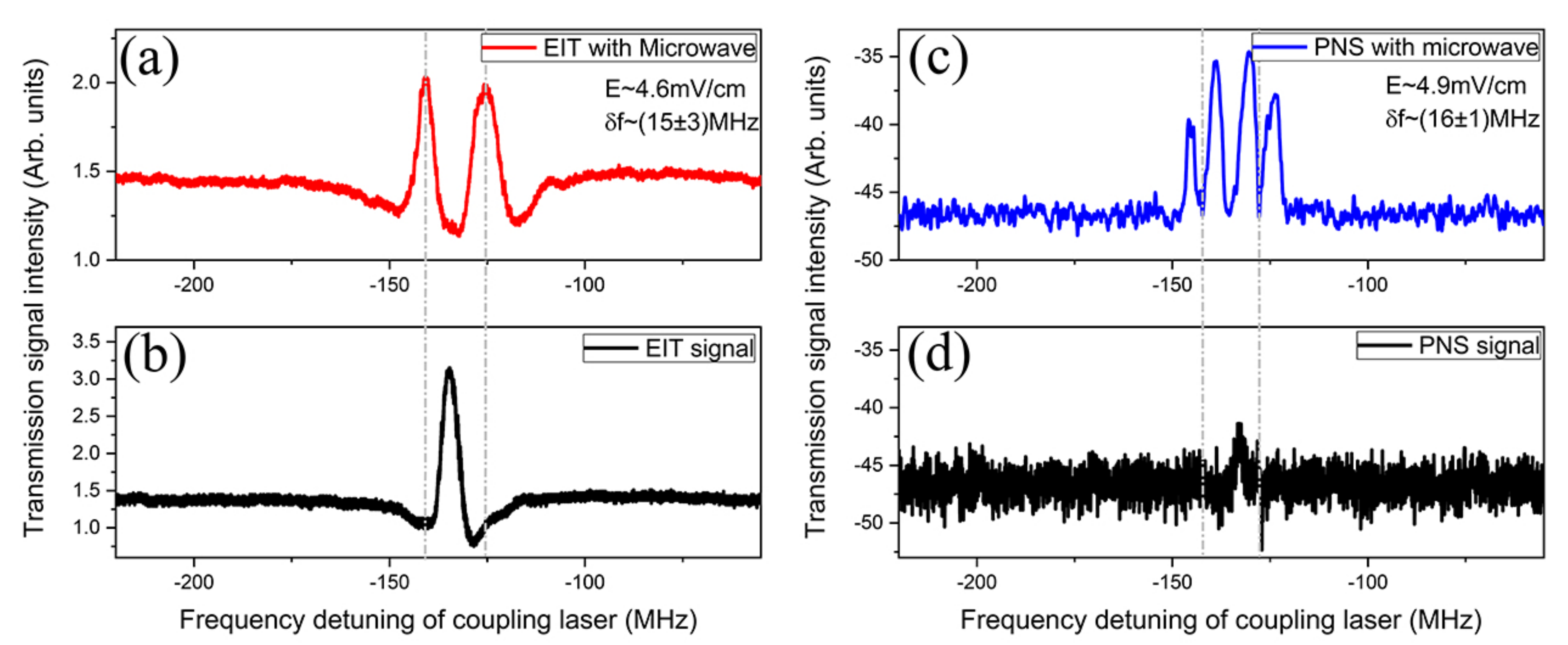}
\caption{(a, b) EIT and (c, d) PNS signals with phase modulated MW fields. The measured frequency splitting is approximately$\delta f=(15\pm 1)MHz$ in EIT spectrum and $\delta f=(16\pm 1)MHz$ in PNS spectrum, respectively, where the corresponding electric field amplitudes are approximately 4.6 mV/cm and 4.9 mV/cm, respectively. The MW frequency is 9.2 GHz, and the phase modulation depth is approximately 20\%.}
\end{figure*}

We simulated the PNS produced under a phase-modulated MW field. A radio-frequency synthesizer was used to generate a 9.2 GHz MW signal with phase modulation depth of approximately 20\%. An antenna was used to transform the MW signal from the signal generator into a space-propagating electromagnetic wave. Typical traces of the response signals are shown in Fig. 7. For the EIT spectrum, the measured frequency splitting is approximately $\delta f=(15\pm 3)MHz$, for which the corresponding electric field amplitude is approximately 4.6 mV/cm. When there is no MW field or phase modulation, the PNS signal is much weaker. The PNS with phase modulated MW fields improves the spectral SNR. The measured frequency splitting in the PNS is approximately $\delta f=(16\pm 1)MHz$, for which the corresponding electric field amplitude is approximately 4.9 mV/cm. Compared with the EIT approaches, the PNS approach presents the same sensitivity with lower uncertainty. In practice, the signal intensities of the modulated PNS are very strong, with values of more than 10 dBm in power, and the SNR of the modulated PNS is typically a factor of 30 greater than the PNS without modulation.

The atoms in the Rydberg state are sensitive to electric and magnetic fields, resulting energy shifts play the same roles in the EIT spectroscopy and PNS spectroscopy. The PNS converts phase noise into amplitude noise via absorption and dispersion effects in the EIT medium. It is realistic to expect that the AT splitting lines have the same shifts for the EIT spectrum and the PNS. The asymmetry of AT spectrum mainly arises from the inhomogeneous distribution of the MW field in the vapor cell. The differences between EIT and PNS spectrum arise from spectral broadening. For the PNS, the two-photon resonance transition converts the phase fluctuations of the coupling field into an intensity fluctuation, the spectral profile is sensitive to two-photon frequency detuning and insensitive to the intensity of the driving laser beams, as shown in Fig. 1(c). The latter represents a small deviation compared to that for the EIT spectrum arising from power broadening of laser beams. The spectral broadening may also arise from a Stark shift caused by auto-ionization of the Rydberg atoms and inhomogeneous broadening in a hot atomic ensemble, which are both sensitive to single-photon detuning [10, 22-26]. 

\section{SUMMARY AND OUTLOOK}
Not all noise in experimental measurements is unwelcome. Certain fundamental noise sources contain valuable information, such as that used in spin noise spectroscopy [27]. The fluctuation dissipation theorem states that the response of a system to a perturbation may be described using the spectrum of the fluctuation exhibited by that system [28]. In principle, the PNS approach is a noise measurement method. When the bandwidth of the modulating frequency is much smaller than the stochastic phase noise bandwidth, the SNR may be improved by narrow-band filtering. 

In conclusion, we have experimentally verified a PNS spectroscopy technique in a Rydberg atom ensemble. Phase fluctuations of the coupling fields cause significant changes in the positioning of two-photon or multiphoton resonances, which means that these fluctuations can be used to realize phase locking of cascading coupling external fields, including those of the laser beam and MW field. We have also proposed a phase-modulated PNS technique, which extend the operating frequency of spectral techniques into the megahertz band. At these measurement frequencies, the noise of the laser system is near the shot noise limit. When technical noise is suppressed within the shot noise limit frequency band, the PNS may represent a promising application for quantum measurements within the MW band.

\begin{acknowledgments}
The project was supported by the National Natural Science Foundation of China (Grants. 61875111, 11974226, and 11774210), the National Basic Research Program of China (Grant 2017YFA0304502), and the 1331Project for Key Subject Construction of Shanxi Province, China.
\end{acknowledgments}

~\\~\\~\\~\\~\\~\\~\\~\\~\\~\\~\\~\\~\\~\\~\\~\\~\\~\\~\\~\\~\\~\\~\\~\\

\end{document}